\documentstyle[12pt]{article}
\begin{document}
\newcommand{\Om}{\Omega}
\newcommand{\df}{\stackrel{\rm def}{=}}
\newcommand{\co}{{\scriptstyle \circ}}
\newcommand{\de}{\delta}
\newcommand{\lb}{\lbrack}
\newcommand{\rb}{\rbrack}
\newcommand{\rn}[1]{\romannumeral #1}
\newcommand{\msc}[1]{\mbox{\scriptsize #1}}
\newcommand{\dsp}{\displaystyle}
\newcommand{\scs}[1]{{\scriptstyle #1}}

\newcommand{\ket}[1]{| #1 \rangle}
\newcommand{\bra}[1]{| #1 \langle}
\newcommand{\vac}{| \mbox{vac} \rangle }

\newcommand{\e}{\mbox{{\bf e}}}
\newcommand{\va}{\mbox{{\bf a}}}
\newcommand{\bc}{\mbox{{\bf C}}}
\newcommand{\br}{\mbox{{\bf R}}}
\newcommand{\bz}{\mbox{{\bf Z}}}
\newcommand{\bq}{\mbox{{\bf Q}}}
\newcommand{\bn}{\mbox{{\bf N}}}
\newcommand {\eqn}[1]{(\ref{#1})}

\newcommand{\cp}{\mbox{{\bf P}}^1}
\newcommand{\n}{\mbox{{\bf n}}}
\newcommand{\sbz}{\msc{{\bf Z}}}
\newcommand{\sn}{\msc{{\bf n}}}

\newcommand{\be}{\begin{equation}}\newcommand{\ee}{\end{equation}}
\newcommand{\bea}{\begin{eqnarray}} \newcommand{\eea}{\end{eqnarray}}
\newcommand{\ba}[1]{\begin{array}{#1}} \newcommand{\ea}{\end{array}}

\newcommand{\cleqn}{\setcounter{equation}{0}}
\makeatletter
\@addtoreset{equation}{section}
\def\theequation{\thesection.\arabic{equation}}
\makeatother

\begin{flushright}
UT-832\\
December, 1998\\
\end{flushright}

\bigskip

\begin{center}
\noindent{\Large \bf Hilbert Space of Space-time SCFT } 
\vskip 3.5mm
{\Large  \bf in $AdS_3$ Supersting }
\vskip 3.5mm
{\Large  \bf and $T^{4kp}/S_{kp}$ SCFT} 

\bigskip
\bigskip

\noindent{\it \large Kazuo Hosomichi and Yuji Sugawara} \\
{\sf hosomiti@hep-th.phys.s.u-tokyo.ac.jp~,~
sugawara@hep-th.phys.s.u-tokyo.ac.jp}
\bigskip

{\it Department of Physics, Faculty of Science, \\
\medskip
          University of Tokyo,\\
\medskip
     Tokyo 113, Japan}
\bigskip
\bigskip

\end{center}
\begin{abstract}
We explore the superstring theory on $AdS_3 \times S^3 \times T^4$
in the framework given in \cite{9806194}. 
We argue on the Hilbert space of ``space-time CFT", and especially
construct a suitable vacuum of this CFT 
from the physical degrees of freedom
of the superstring theory in bulk.
We first construct  it explicitly in the case of $p=1$, and then present
a proposal for the general cases of $p>1$.

After giving some completion of the GKS's constructions
of  the higher mode operators (in particular, of those    
including  spin fields),
we also make  some comparison   between the space-time CFT and 
$T^{4kp}/S_{kp}$ SCFT, namely, with respect to 
the physical spectrum of chiral primaries and 
some algebraic structures of bosonic and fermionic oscillators 
in both theories.
We also observe how our proposal about the Hilbert space of space-time
CFT leads to a satisfactory correspondence between the spectrum of
chiral primaries of both theories in the cases $p>1$.

\end{abstract}

\newpage

\section{Introduction}
\cleqn

In the recent studies of the proposed duality between 
the supergravity  theory  (SUGRA) on $AdS_{d+1} \times K $ and 
the $d$-dimensional  conformal field theory (CFT) 
defined on the boundary of $AdS_{d+1}$
\cite{9711200,9802150,9802109}, the works
on $AdS_3$ have got a special status. 
Both of the $AdS_3$-gravity and 2-dimensional CFT have infinite-dimensional
local symmetries\cite{Brown}, which bring us a high ability of calculation. 
Thanks to this fact, several fruitful results 
related with the black-hole physics are given
\cite{9712251,9802076,9804085,9804111,9804166,9805165,9806026}. 
Among other things, it is remarkable that 
 we can expect rich possibilities 
to establish the $AdS/CFT$-duality in 
{\em stringy} level. 
As is well-known,  a classical solution of SUGRA 
corresponds to a 2-dimensional CFT on the world-sheet of first quantized 
superstring. Therefore the $AdS_3/CFT_2$-duality in stringy level naturally
leads to the correspondence between the two 2-dimensional CFTs - the ``
world-sheet CFT" and the ``space-time CFT"\footnote
            {In many litarature, the terminology ``boundary CFT"
            is often used in this sense. 
            Throughout this paper, we shall use the terminology 
            ``space-time CFT" in the meaning of the one 
            proposed in \cite{9806194} as a candidate for the boundary 
            CFT.}, 
the latter of which is defiend on the boundary of 
$AdS$ space. Among other things
Giveon, Kutasov and Seiberg (GKS) have obtained a 
remarkable result \cite{9806194}: 
The  typical operators 
(the generators of superconformal algebra) in the space-time CFT  have been 
given directly from the physical degrees of freedom of the world-sheet CFT. 
Recently,
there are some further developments \cite{9811002,9811245,9812027,9812046} 
along this line.

However, there still exists an important question which is not yet answered:
What is the Hilbert space of space-time CFT? Especially, what is the suitable
vacuum on which the GKS's operators in space-time CFT should act?
Needless to say, answering this question  
is  significant  for
calculating  correlators in the  space-time CFT  (and of course, 
according to the standard programs in recent $AdS$-business, 
comparing them with those 
calculated  by SUGRA or string theory in bulk...).
Solving this problem  is  one of the main purposes of this paper. 
We will propose an explicit realization of the  ``space-time vacuum".

Another important subject which may be interested  by many 
theoretists studying $AdS/CFT$-duality is to confirm the equivalence
(or some relationship) between the space-time CFT of GKS and 
the SCFT on the symmetric orbifold $T^{4kp}/S_{kp}$, which is derived 
by some brane dynamics. We carefully investigate the spectrum of 
chiral primary fields in space-time CFT, and discuss    
the equivalence with that of $T^{4kp}/S_{kp}$-SCFT.

~

This paper is organized as follows:
We shall begin in section 2 by fixing our convetion and 
giving a short review of 
the formulation of GKS for convenience of readers.
We will  also  present  some non-trivial completion of the calculation
of GKS (all of the explicit forms  of the higher modes of supercurrents). 

In section 3, we discuss the desired vacuum of space-time CFT. 
We first construct it  explicitly for the case of $p=1$, 
($p$ is equal to the NS1 charge in the set up in 
\cite{9806194}.) and then we propose a candidate of the Hilbert space
with a suitable vacuum for the general case of $p>1$. 

In section 4,  in order to study the relation between the space-time CFT and 
$T^{4kp}/S_{kp}$-SCFT,
we start by working out all  the  bosonic and fermionic oscillators  
along $T^4$ in the space-time CFT
that acts properly on the vacuum defined in section 3.
Making use of this knowledge, we construct all of the mode operators 
of typical chiral primary fields, and discuss about the equivalence 
with the physical spectrum of $T^{4kp}/S_{kp}$-SCFT.
Especially, we will show how our proposal of the Hilbert space can 
resolve the superficial contradictions about the chiral primary 
spectrum for the general case $p>1$.
 We also comment on 
the relation  between the SCA given by GKS and 
another SCA which is  constructed 
as the quadratic forms 
of the bosonic and fermionic oscillators along $T^4$.

Section 5 is devoted to give some comments on several open problems.

~

\section{Superstring on $AdS_3\times S^3 \times T^4$}
\cleqn

  Type IIB superstring theory on the spacetime $AdS_3\times S^3\times T^4$
with NS1 and NS5-brane charges was investigated 
in \cite{9806194}.
  One of the main results there was that one can construct 
an $N=4$ SCA, which should act on the boundary of $AdS_3$,
out of certain physical operators in the world-sheet theory.
  This SCA is considered to be identified with the one
discovered by Brown and
Henneaux\cite{Brown}, which emerges when one considers the
(super)gravity theory on a three-dimensional space-time with a negative 
cosmological constant.
This aspect is also realized elegantly
from the viewpoint of the 3-dimensional Chern-Simons gravity
\cite{9802076,9804111}.

Let us give a brief review of the work \cite{9806194} here.
There is a well-known solution in type IIB supergravity which is identified
with the bound state of some NS1 and NS5-branes\cite{CHS,DGHR,9601177}.
  If we approach the near-horizon region of this solution, the geometry
reduces to that of $AdS_3\times S^3\times {\cal M}^4$, where ${\cal M}^4$
is some four-dimensional spatial manifold.
  The $AdS_3$ part has the following metric:
\begin{equation}
  ds^2= l^2(d\varphi^2 + e^{2\varphi}d\gamma d\bar{\gamma})\;\; ; \;\;
  l^2 = l_s^2 k^\prime
\end{equation}
where $l_s$ is the string length and $k^\prime$ is an integer which is roughly
regarded as the number of NS5-branes.
  Taking  the presence of NS-NS B-field into account, one can write down the
action of the string on $AdS_3$-space.  
After evaluating some quantum corrections it becomes as
\begin{equation}
  {\cal L} =
  \partial\varphi\bar{\partial}\varphi 
  -\frac{2}{\alpha_+}\hat{R}^{(2)}\varphi
  +\beta\bar{\partial}\gamma + \bar{\beta}\partial\bar{\gamma}
  -\beta\bar{\beta}\exp\left(-\frac{2\varphi}{\alpha_+}\right)
\end{equation}
where $\alpha_+= \sqrt{2k^\prime-4}$.
  The worldsheet theory has two copies of $SL(2,R)$ current algebra of level
$k^\prime$ (the left mover and right mover). 
These are represented in the standard form of Wakimoto construction
\cite{Wakimoto}: 
\begin{eqnarray}
  j^- &=& \beta  \nonumber \\
  j^3 &=& \beta\gamma + \frac{\alpha_+}{2}\partial\varphi \nonumber  \\
  j^+ &=& \beta\gamma^2 + \alpha_+\gamma\partial\varphi 
          + k^\prime \partial\gamma
\end{eqnarray}
together with the OPE's of the fields
\begin{equation}
  \varphi(z)\varphi(w) \sim -\ln (z-w) \;\; ; \;\;
  \beta(z)\gamma(w) \sim \frac{1}{z-w}
\end{equation}
Of course, the right mover can be written in the same form.

  One can apply, at least formally, the same argument to the $S^3$ part,
which yields the following expression for $SU(2)$ current algebra of level
$k^{\prime\prime}$:
\begin{eqnarray}
  k^- &=& \tilde{\beta}  \nonumber \\
  k^3 &=& \tilde{\beta}\tilde{\gamma} 
          + \frac{i\alpha_+}{2}\partial\tilde{\varphi} \nonumber  \\
  k^+ &=& -\tilde{\beta}\tilde{\gamma}^2
          -i\alpha_+\tilde{\gamma}\partial\tilde{\varphi}
          +k^{\prime\prime}\partial\tilde{\gamma}
\end{eqnarray}
where $\alpha_+= \sqrt{2k^{\prime\prime}+4}$, together with the following
OPE's
\begin{equation}
  \tilde{\varphi}(z)\tilde{\varphi}(w) \sim -\ln (z-w) \;\; ; \;\;
  \tilde{\beta}(z)\tilde{\gamma}(w) \sim \frac{1}{z-w}
\end{equation}
  The constant $\alpha_+$ in the above must be the same as the one in the
$AdS_3$ part.
  It may be convenient to note the relation;
\begin{equation}
  \frac{\alpha_+^2}{2}= k'-2 = k^{\prime\prime}+2 \equiv k .
\end{equation}

  To supersymmetrize the worldsheet theory one only has to add some free
fermions $\psi^A, \chi^a$ which have the following OPE's\footnote
   {Although our convention about the signature of metric $\eta^{AB}$
     is the same as the one given in \cite{9806194}, it is more
     natural in a physical reason to take the inverse sign for it.
    In this case, we must suppose that the bosonic $SL(2,\br)$-current 
     should have the level $-k'$, and the fermionic $SL(2,\br)$-current 
    should have the level $+2$. We should thank I. Bars for his comment
    on this fact. }
\begin{eqnarray}
  \psi^A(z)\psi^B(w) &\sim& \frac{\eta^{AB}}{z-w}\;\;\;\; ; \;\;\;\;
  \eta^{AB} = {\rm diag} (++-) \\
  \chi^a(z)\chi^b(w) &\sim& \frac{\delta^{ab}}{z-w}
\end{eqnarray}
to the bosonic theory.
  Using these fermions one can construct $SL(2,R)$ and $SU(2)$ currents
of levels $-2$ and $2$, respectively,
\begin{eqnarray}
  j_\psi^\pm = \pm \psi^\pm \psi^3 \;\;,\;\;
  j_\psi^3   = \frac{1}{2}\psi^+\psi^- &;&
  \psi^\pm   = \psi^1 \pm i\psi^2 \\
  k_\chi^\pm = \mp \chi^\pm \chi^3 \;\;,\;\;
  k_\chi^3   = \frac{1}{2}\chi^+\chi^- &;&
  \chi^\pm   = \chi^1 \pm i\chi^2
\end{eqnarray}
  The total currents are defined as
\begin{equation}
  J^A= j^A+j_\psi^A \;\;,\;\;
  K^a= k^a+k_\chi^a
\end{equation}
both of which are of level $k= k^\prime-2 = k^{\prime\prime}+2$.
  Furthermore we introduce four bosons $Y^i$ and four fermions $\lambda^i$
for $T^4$ degrees of freedom.
  The worldsheet theory has the following energy-momentum tensor $T(z)$
and the supercurrent $G(z)$:
\begin{eqnarray}
  T(z) &=&
   \frac{1}{k}\left(j^Aj_A+k^ak^a\right)
  -\frac{1}{2}\left(
   \psi^A\partial\psi_A+\chi^a\partial\chi^a
  +\partial Y^i\partial Y^i+\lambda^i\partial\lambda^i\right) \\
  G(z) &=&
   \sqrt{\frac{2}{k}}\left(
   \psi^Aj_A-\frac{i}{6}\epsilon^{ABC}\psi_A\psi_B\psi_C
  +\chi^ak^a-\frac{i}{6}\epsilon^{abc}\chi^a\chi^b\chi^c\right)
  +i\lambda^i\partial Y^i
\end{eqnarray}

  In \cite{9806194} the ``space-time'' $N=4$ SCA was constructed out of vertex
operators in the world-sheet theory.
  The Virasoro operators ${\cal L}_n$ and the mode operators of $SU(2)$ current
${\cal K}_n^a$ have the following form:
\begin{eqnarray}
  {\cal L}_n &=&
  -\sqrt{\frac{k}{2}}\oint dz e^{-\phi}
  [(1-n^2)\psi^3\gamma^n+\frac{n(n-1)}{2}\psi^-\gamma^{n+1}
   +\frac{n(n+1)}{2}\psi^+\gamma^{n-1}] \\
  {\cal K}_n^a &=&
  \sqrt{\frac{k}{2}}\oint dz e^{-\phi}\chi^a\gamma^n
\end{eqnarray}
where $\phi$ is the bosonized super-reparametrization ghost.
  In the above expressions we are working in the
$(-1)$-picture and we can obtain the expression 
in the 0-picture by the standard picture-changing operation 
\cite{FMS}.
  The fermionic generators are constructed in terms of the spin fields of the
worldsheet theory.
  In order to obtain the explicit forms of  these generators we first give
the bosonization formula of fermions:
\begin{eqnarray}
 \sqrt{2} \, e^{\pm iH_1} &\equiv& \psi^{\pm 1} 
 \equiv \psi^1\pm i\psi^2 \nonumber \\
 \sqrt{2} \, e^{\pm iH_2} &\equiv& \psi^{\pm 2} 
 \equiv \chi^1\pm i\chi^2 \nonumber \\
 \sqrt{2} \, e^{\pm iH_3} &\equiv& \psi^{\pm 3} 
 \equiv \chi^3 \pm \psi^3 \nonumber \\
 \sqrt{2} \, e^{\pm iH_4} &\equiv& \psi^{\pm 4} 
\equiv \lambda^1 \pm i\lambda^2  \nonumber\\
\sqrt{2} \,  e^{\pm iH_5} &\equiv& \psi^{\pm 5} 
\equiv \lambda^3 \pm i\lambda^4
\end{eqnarray}
up to some cocycle factors that ensure all the fermions to be mutually
anti-commuting.
  Then we define  the spin fields  
$S^{\epsilon_1\epsilon_2\epsilon_3\epsilon_4\epsilon_5}(z)=S^A(z)$, where
$\epsilon_I$ takes values $\pm1$,  as
\begin{equation}
  S^A(z) \equiv \exp( \frac{i}{2}\sum_I \epsilon_IH_I(z) )
\end{equation}
up to cocycle factors.
  The calculations including  cocycle factors are 
rather complicated, and are not discussed here. One only has to
remember that it is defined consistently with the OPE's of fermions and
spin fields.
  We summarize the relevant formulas in the appendix.

  The fermionic generators in SCA can be  written in terms of spin fields.
Although  only the explicit forms of the ``zero-mode" part of 
these (correspondign to the global space-time SUSY) are given in 
\cite{9806194}, we can now write down the general higher
modes of supercurrents:   
\begin{eqnarray}
  {\cal G}_r^{+\epsilon} &\equiv&
  \oint dz e^{-\frac{\phi}{2}}[
   (r+\frac{1}{2})\gamma^{r-\frac{1}{2}}S^{++-\epsilon\epsilon}(z)
  -(r-\frac{1}{2})\gamma^{r+\frac{1}{2}}S^{-++\epsilon\epsilon}(z)] 
\nonumber \\
  {\cal G}_r^{-\epsilon} &\equiv&
  \oint dz e^{-\frac{\phi}{2}}[
   (r+\frac{1}{2})\gamma^{r-\frac{1}{2}}S^{+-+\epsilon\epsilon}(z)
  -(r-\frac{1}{2})\gamma^{r+\frac{1}{2}}S^{---\epsilon\epsilon}(z)]
\label{stg}
\end{eqnarray}
In particular, for $r=\pm 1/2$ we have
\begin{eqnarray}
  {\cal G}_{\frac{1}{2}}^{+\epsilon} \equiv
  \oint dz e^{-\frac{\phi}{2}}S^{++-\epsilon\epsilon}(z) &,&
  {\cal G}_{-\frac{1}{2}}^{+\epsilon} \equiv
  \oint dz e^{-\frac{\phi}{2}}S^{-++\epsilon\epsilon}(z) \nonumber \\
  {\cal G}_{\frac{1}{2}}^{-\epsilon} \equiv
  \oint dz e^{-\frac{\phi}{2}}S^{+-+\epsilon\epsilon}(z) &,&
  {\cal G}_{-\frac{1}{2}}^{-\epsilon} \equiv
  \oint dz e^{-\frac{\phi}{2}}S^{---\epsilon\epsilon}(z),  
\end{eqnarray}
which, of course, are the same as  the original result 
given in \cite{9806194}.  

Recently, the similar expressions for the supercurrents based on 
some affine Lie superalgebra were given\cite{9811002}. It may be interesting 
to study the relation between them and our results here \eqn{stg}.

  Note that we are restricting ourselves to the spin fields of definite
chirality, namely, $\Pi \epsilon_I = -1$, due to the requirement of mutual
locality of supercharges.
  There is a further restriction coming from the BRST invariance: for
example, of all the operators which have the form
\[ \oint dz e^{-\frac{\phi}{2}} S^A \]
only those with $\epsilon_1\epsilon_2\epsilon_3 = -1$ are BRST
invariant.\footnote{
We work in a convention in which the definition of the third of the five
signs is opposite to that of ref. \cite{9806194}. }

After some straightforward but lengthy calculations, we can directly 
show that the operators $\{{\cal L}_n, \, {\cal K}^a_n , \, 
{\cal G}^{\alpha A}\}$
actually generate the correct $N=4$ SCA with central charge $c=6pk$, 
where we impose the constraints
$\dsp \oint dz \gamma^{-1}\partial\gamma =p$. 
As was discussed in \cite{9806194},
this condition is the essential part to reproduce the correct central term 
in SCA.
  
There is one subtlety which might lead one to a misunderstanding.
This condition might sound  peculiar from 
the viewpoints of usual string theory. In fact, as was claimed 
in the recent paper  \cite{9812046}, it is reasonable to think that
$\dsp \oint dz \gamma^{-1}\partial\gamma = 0$ is realized 
on the correct string vacuum. 
In \cite{9812046} it was further argued  that 
the correct central term in the   Virasoro algebra of the boundary CFT
should arise from
the contribution of non-connected world-sheets, not from
$\dsp \oint dz \gamma^{-1}\partial\gamma =p$.
At first sight this argument might seem to contradict with the treatment of
GKS. But this is not the case. One should consider that the equality 
\be
 \oint dz \gamma^{-1}\partial\gamma \, \ket{0}= 0 ,
\ee
holds on the correct vacuum of string theory on $AdS_3$-background, and 
at the same time, should suppose that 
the condition 
\be
 \oint dz \gamma^{-1}\partial\gamma \, \vac = p\vac  
\ee
gives  the {\em definition\/} 
of suitable vacuum $\vac$ in  the space-time CFT of GKS
(which will be called as ``space-time vacuum'' in this paper).
Of course $\vac$ need  not (and should not) be equal to
$\ket{0}$. Although the space-time CFT  
is defined by the degrees of freedom
on  world-sheet, {\em it is not the $AdS_3$-string theory itself\/}, 
and we have no contradiction if we take a different vacuum $\vac$
for this theory.

In the last section we will again give some comments to clarify 
the compatibility between the GKS's construction of space-time CFT
and the work \cite{9812046}.

~

\section{Vacuum of the Spacetime CFT}
\cleqn

\subsection{Space-time Vacuum with $p=1$}

It is an important problem to find out the vacuum vector of the space-time
CFT. Clearly, we must find it in the physical Hilbert space 
of string theory,  and it should have 
the superconformal invariance with respect
to the superconformal algebra of the  space-time CFT.
Moreover, it should be emphasized that it  is indeed different from 
the usual vacuum of superstring theory, as we already commented 
at the last of previous section. 
This is because the condition
$\dsp \oint dz \gamma^{-1}\partial\gamma  = p$
must be realized  on this vacuum. 

Namely,
we shall impose on the vacuum $\left|{\rm vac}\right>$ the
following conditions:
\begin{enumerate}
  \item $\left|{\rm vac}\right>$ is  BRST invariant in ths sense of 
            superstring theory,
      \be 
        Q_{\msc{BRST}} \vac =0  \label{brst}.
        \ee
  \item $\left|{\rm vac}\right>$ is primary and 
             has the global superconformal invariance in the sense of 
     space-time CFT (up to some BRST-exact terms).
          \begin{eqnarray}
      {\cal L}_n\left|{\rm vac}\right>             = 0 &;&(n\ge -1) \nonumber\\
      {\cal G}^{\alpha A}_r \left|{\rm vac}\right> = 0 &;&(r\ge -\frac{1}{2})
                                                             \nonumber \\
      {\cal K}^a_n\left|{\rm vac}\right>           = 0 &;& (n\ge 0 )
                                                      \label{sci}
    \end{eqnarray}
  \item The vacuum has the winding number $p$ of $\gamma$.
    \begin{equation}
      \oint dz \gamma^{-1}\partial\gamma \left|{\rm vac}\right> =
      p\left|{\rm vac}\right>  \label{winding gamma}
    \end{equation}
\end{enumerate}
 
Let us start with  a natural ansatz of  the form 
$\left|{\rm vac}\right> = c V(0) \left|0\right>$, where 
$c$ denotes the reparametrization ghost (spin $-1$) and $V(z)$ should
be some primary field with conformal weight 1, which will be constructed 
below. 

To this aim (in particular, to realize the third condition
\eqn{winding gamma}) it is useful to introduce the following scalar
fields $X$, $Y$ bosonizing the $\beta \gamma$-system 
(of $AdS_3$ part of worldsheet theory);
as
\begin{eqnarray}
  \beta &=& i\partial Y e^{-Z} \\
  \gamma&=& e^Z \\
  Z     &=& X+iY; \;\;\;\;\;
  X(z)X(w)\sim Y(z)Y(w) \sim -\ln (z-w)
\end{eqnarray}
We can easily obtain $\gamma^{-1}\partial\gamma = \partial Z$.
By this relation, it seems plausible to 
take the vertex operators such as $\sim e^{ipY}$ or $\sim e^{-pX}$ 
as candidates for $V(z)$. Unfortunately, the requirements of
the BRST invariance  \eqn{brst} and superconformal invariance in
space-time theory \eqn{sci} are very strong 
conditions, and hence it seems that almost all possibilties are
excluded in the cases of generic $p$. 
The best we can do here  is to set $p=1$ for the time being.
We will later discuss the cases with generic $p$. 

Now, we can easily find that the operator $e^{iY}$ is actually a 
primary field with conformal weight 1
under the correct background charge fixed by that of
the $U(1)$-current $\beta\gamma$. 
Also we can check that 
this commutes with all  the zero-modes of  bosonic $SL(2,\br)$ 
currents up to total derivative terms\footnote
   {There is only  one non-trivial OPE;
     $$j^- (z) e^{iY}(0) \sim \frac{1}{z^2}e^{-X}(0) 
       + \frac{1}{z}\partial (e^{-X}) (0). $$
    The existence of the term with $\frac{1}{z^2}$ singularity in this OPE
    is crucial for our arguments  with respect to the BRST invariance
     given below.}(It is obvious that this commutes with all of
      the fermionic currents and the bosonic $SU(2)$-currents.)
It immediately gives the  superconformal invariance \eqn{sci}. 
The condition of BRST invariance is more non-trivial.
As we already know this candidate has a  suitable conformal weight, 
we must only confirm the invariance under the fermionic part of 
BRST-transformation (including the supercurrent $G(z)$ in the worldsheet). 
Unfortunately, we find that this is not actually BRST-invariant.
But  a careful analysis  gives us the following completion possessing 
the full BRST-invariance:
\be
\vac = c e^{iY}(0) \ket{0} + \frac{1}{\sqrt{2k}} 
\eta \, e^{\phi} e^{-X} \psi^+ \ket{0}.
\label{vac}
\ee
In the above expression, 
$\eta(z)$ denotes a fermionic field with spin 1 composing 
the bosonized super-ghost system 
together with the fermionic partner
$\xi$ (spin 0) and the bosonic field $\phi$ already introduced 
in the previous section \cite{FMS}.
One can show that this state 
satisfies all of the above conditions for the ``space-time vacuum''
\eqn{brst}, \eqn{sci}, \eqn{winding gamma}, 
and is no other than
our proposal of space-time vacuum in the case of $p=1$.

There is one comment which may be useful for some explicit calculations:  
The second term in the RHS of \eqn{vac} vanishes in almost all
cases when some vertex operators with  a negative 
picture (for example, operators 
in the $(-1)$-picture for NS sector, and in the $(-1/2)$-picture 
for R sector, which are ``standard" pictures we often use) act on it, 
although we need this term to ensure the BRST-invariance.  

\subsection{The Space-time Vacuum with General $p$}

Now, how can we construct the vacuum for the general cases with $p >1$?

To answer this question, we would like to make the following proposal;
{\em We should identify the Hilbert space of space-time CFT
with (at least, a subspace of) the physical Hilbert space 
of $p$-string states in second quantized superstring theory. 
In other words, we should consider $p$-copies of 
the world-sheet of superstring 
to define the space-time CFT.} In this set up
we can construct a suitable vacuum in the following manner
\be
\vac \df \frac{1}{p!}\sum_{\sigma \in S_p}
 \bigotimes_{i=1}^p \, \vac^{(\sigma(i))}  ~ \in
 \left[ ( {\cal H}_{\msc{phys}})^{\otimes p}\right]^{S_p}  ,
\label{vac p}
\ee
where  each $\vac^{(i)}$ is defined as \eqn{vac} for each world-sheet,
and ${\cal H}_{\msc{phys}}$ 
stands for the physical Hilbert space of single string.
The superscript $S_p$ expresses the $S_p$-invariant subspace.

General operators in space-time CFT (generators of SCA, and some primary
operators, and so on) should act on this Hilbert space in a natural way;
$\dsp {\cal A} \equiv \sum_i 1\otimes 1 \otimes 
\cdots \otimes {\cal A}^{(i)} \otimes \cdots \otimes 1 $, 
where ${\cal A} $ denotes an arbitrary operator in space-time CFT and 
${\cal A}^{(i)} $ acts only on the the $i$-th factor of 
$({\cal H}_{\msc{phys}})^{\otimes p}$. 
It is obvious that \eqn{vac p} has all of the desired properties.
Especially, we can obtain the correct winding number of $\gamma$ ;
\be
\oint dz \gamma^{-1}\partial\gamma \left|{\rm vac}\right> =
      \sum_{i=1}^p \, 1 \vac  = p\left|{\rm vac}\right> ,  
\ee
and of course the correct central charge $\dsp c= \sum_{i=1}^p \, c^{(i)} \equiv 6pk $. 
 
One might feel this ansatz for the Hilbert space of the space-time CFT 
as somewhat artificial. One must, however, remember  that, in the standard 
argument of $AdS/CFT$-correspondence, the Hilbert space of boundary CFT
should include multi-particle states of bulk-theory.
Moreover, we can show the following rough estimation of physical degrees 
of freedom  which supports our above proposal on the  Hilbert space:
As the representation theory of current algebra, the world-sheet CFT
has level $\sim k$, and so we should have an unitarity bound for $SU(2)$ 
(or $SL(2, \br)$) charge of the order $k$ \cite{9806024}. 
On the other hand, 
the space-time CFT should have the unitarity bound for $R$-charge of 
the order $pk$, which is known as the name 
``stringy exclusion principle" \cite{9804085}.
In this way, one can find that the space-time CFT includes roughly 
$p$-times larger degrees of freedom than those on a {\em single} 
world-sheet. One of the most simple (and perhaps, natural)
solutions to fill this gap of degrees of freedom is no other than  
introducing  $p$-strings Hilbert space!

If one accept the  above proposal, 
that is, {\em ``space-time CFT should describe 
the second quantized superstring theory"}, it may be more natural 
to consider the Matrix string theory proposed
in \cite{9703030} instead of the ``old-fashoned"
fundamental string.
From this viewpoint it may be important 
 to consider general ``twisted sectors".
(On the other hand, the simplest construction of Hilbert space  
given above corresponds to the ``untwisted sector".)
To explain this, let us consider the following setup:
Let $(n_1, \ldots , n_l)$ be an arbitrary  partition of $p$,
Namely, $n_1 \geq n_2 \geq \cdots \geq  n_l >0$ 
with $\dsp \sum_{i=1}^l n_i =p$.
Then, we  can introduce the single-string Hilbert space for
the ``$\bz_{n_i}$-twisted string"  ${\cal H}^{(n_i)}_{\msc{phys}} 
(\cong {\cal H}_{\msc{phys}}^{\sbz_{n_i}})$
for each $n_i$, which is  defined  by the ``screwing procedure", 
something like
\be
X_{(i)}^{\mu}(e^{2\pi i} z) = X_{(i+1)}^{\mu}(z) ,
\ee
and by imposing $\bz_{n_i}$-invariance.

In this setup, our space-time vacuum should be constructed in 
the Hilbert space $\dsp \left[
\bigotimes_{i=1}^l {\cal H_{\msc{phys}}}^{(n_i)} \right]
^{\Gamma_{n_1,\ldots, n_l}}$, where $\Gamma_{n_1,\ldots, n_l}$
means the symmetry group composed of arbitrary 
permutations among  twisted strings of the same ``length".

However, if one tries to construct our space-time vacuum naively
for each world-sheet of $\bz_{n_i}$-twisted string, one might feel that   
our above discussion leads to a contradiction.
Since we now have only $l(<p)$ worldsheets, it seems that the total
central charge of the space-time CFT becomes
$\dsp c=\sum_{i=1}^l c^{(i)} =6lk \neq 6pk$.
We should here again enphasize that  in each world-sheet of $l$ strings, 
our space-time vacuum $\vac^{(i)}$ should have the property:
$\dsp \oint dz \gamma^{-1}\partial\gamma \left|{\rm vac}\right>^{(i)} =
\left|{\rm vac}\right>^{(i)} $. We know no solution possessing 
the winding number $p>1$ {\em which is assigned to a single
worldsheet}
 under the requirements of BRST-invariance
and superconformal invariance of space-time. 
How can we overcome this difficulty?

The simplest possibilty is as follows: {\em we should assign the level 
$n_i k$ current algebras instead of $k$  for each world-sheet 
of $\bz_{n_i}$-twisted string.} If this is indeed the case, 
each worldsheet yields  
the contribution $c^{(i)}=6 n_i k$, and we can obtain
the correct central charge $\dsp c=\sum_{i=1}^l c^{(i)} =6pk $.
Furthermore, we will observe in the next section, how this claim 
about the enhancement of level gives us the suitable spectrum 
of chiral primary operators in the space-time CFT.

In the rest of this section, let us give a heuristic explanation 
about how this enhancement occurs for the world-sheets of twisted 
strings. For a complete understnding  of it, we may have to 
formulate carefully the Matrix string theory on  $AdS_3$-background.
We would like to study this problem  intensively in future work.

Let us consider only the maximally twisted sector to avoid unnecessary
complexity. That is, assume that there is  only  a 
single world-sheet of $p$-joined string
$\Sigma (\cong \cp)$. Since our string is now the ``longest string"
possessing the $p$-times longer length, 
let us  consider the following covering 
\be
\Phi ~ : ~ z  ~\in \Sigma ~ \longmapsto ~  z^p ~ \in  
\hat{\Sigma} (\cong \cp)
\ee
in order to ``normalize" the unit of excitations.

Imagine some current algebra $J^a(z)$ is defined  on $\Sigma$ with level $k$.
We would like to construct 
a proper current $\hat{J}^a(\hat{z})$ defined on
the ``normalized" world-sheet $\hat{\Sigma}$, 
on which we should formulate our space-time CFT,
so that the relation $\Phi^* (\hat{J^a}d\hat{z}) = J^a \, dz$ holds.
However, $J^a(z)$ cannot act suitably on ${\cal H}^{(p)}_{\msc{phys}}
\cong {\cal H_{\msc{phys}}}^{\sbz_p}$, because $J^a(z)\, dz$ does not
possess $\bz_p$-invariance. Or equivalently, one can also say that
$J^a$ does not yield by this simple relation 
the current $\hat{J}^a$ which is {\em 
a single valued 1-form on $\hat{\Sigma}$}

Therefore we must instead adopt 
the follwoing definition of $\hat{J^a}$:
\be
\Phi^*(\hat{J}^a d\hat{z}) = \sum_{n=0}^{p-1} \, \Omega^{(n)*} (J^a dz) ,
\ee
where $ \dsp \Omega^{(n)} ~ : z \, \longmapsto \, e^{2\pi n i /p} z$
is a $\bz_p$-rotation on $\Sigma$.
It is easy to see that this current $\hat{J^a}$ has actually the level
$kp$, which is our claim to be proved.

We also comment on a simple evaluation of the  degrees of freedom.
Consider first the untwisted sector, namely, we have $p$ worldsheets
of ``short" strings, and each worlsheet has a level $k$ current algebra.
If these $p$ strings are  completely joined  one after another 
by interactions of Matrix string theory, we will get only one long string.
The  simplest and maybe the most plausible way to impose the invariance
of degrees of freedom  through this joining process
is to require the level of current algebras on the world-sheet 
of long string should be enhanced to the value $pk$.

~

\section{Comparison with the SUSY $\sigma$-Model on $T^{4kp}/S^{kp}$}
\cleqn

In the well-known arguments of $AdS_3/CFT_2$-duality,
it is belived that the SCFT defined as the $N=(4,4)$ 
supersymmetric sigma model on a symmetric orbifold  $T^{4kp}/S^{kp}$
is one of the most powerful candidates of the boundary CFT. 
In this sense it is an important task to compare the space-time CFT
with this $T^{4kp}/S^{kp}$ SCFT. We already know that 
both of them have $N=(4,4)$ superconformal symmetry 
with the same central charge $c=6pk$. But, in order to establish
the equivalence (or some relation) between them, we must still
clarify the spectrum of these SCFTs, especially, the spectrum 
of chiral primaries. 

In this section, we first explore further the operator algebra
in the space-time CFT. We would like to find
the complete set of the bosonic and fermionic oscillators 
inspired by the $T^{4kp}/S^{kp}$-theory
(the Heisenberg algebras along $T^4$).

Next, we will discuss the issue of  the physical spectrum 
of chiral primaries in the space-time CFT. We will observe
how we can obtain the  spectrum which is equivalent  with that of 
$T^{4kp}/S^{kp}$-SCFT in the Hilbert space of space-time CFT
proposed in the previous section.

\subsection{Construction of Mode Operators Along $T^4$}

  In the $\sigma$-model on $T^{4kp}/S^{kp}$ there is a ``diagonal'' $T^4$
which is free from any operations of orbifolding.
  Therefore the spacetime CFT contains a sector which is nothing but a
free SCFT on $T^4$.
  And we can find rather easily the corresponding degree of freedom in the
worldsheet theory.
 
  Define $\left\{ a^{AK}_n, x^{AK}_n, b^{\alpha K}_r \right\}$ as
\begin{eqnarray}
  ia^{++}_n &=& -\sqrt{\frac{k}{2}}\oint dz e^{-\phi}\psi^{+5}\gamma^n
             \sim -\sqrt{\frac{k}{2}}\oint dz i\partial(Y^3+iY^4)\gamma^n 
      \nonumber \\
  ia^{+-}_n &=& +\sqrt{\frac{k}{2}}\oint dz e^{-\phi}\psi^{+4}\gamma^n
             \sim +\sqrt{\frac{k}{2}}\oint dz i\partial(Y^1+iY^2)\gamma^n 
     \nonumber \\
  ia^{-+}_n &=& +\sqrt{\frac{k}{2}}\oint dz e^{-\phi}\psi^{-4}\gamma^n
             \sim  +\sqrt{\frac{k}{2}}\oint dz i\partial(Y^1-iY^2)\gamma^n 
    \nonumber  \\
  ia^{--}_n &=& +\sqrt{\frac{k}{2}}\oint dz e^{-\phi}\psi^{-5}\gamma^n
             \sim  +\sqrt{\frac{k}{2}}\oint dz i\partial(Y^3-iY^4)\gamma^n ,
\label{a}
\end{eqnarray}
where  readers must understand ``$\sim $" in the sense of equivalence 
by the picture changing. Similarly, we define
\begin{eqnarray}
  x^{++}_n &=& +\frac{1}{2}\oint dz e^{-\phi}(Y^3+iY^4)\gamma^n
  (\gamma\psi^-+\gamma^{-1}\psi^+-2\psi^3) \nonumber  \\
  x^{+-}_n &=& -\frac{1}{2}\oint dz e^{-\phi}(Y^1+iY^2)\gamma^n
  (\gamma\psi^-+\gamma^{-1}\psi^+-2\psi^3)  \nonumber \\
  x^{-+}_n &=& -\frac{1}{2}\oint dz e^{-\phi}(Y^1-iY^2)\gamma^n
  (\gamma\psi^-+\gamma^{-1}\psi^+-2\psi^3)  \nonumber \\
  x^{--}_n &=& -\frac{1}{2}\oint dz e^{-\phi}(Y^3-iY^4)\gamma^n
  (\gamma\psi^-+\gamma^{-1}\psi^+-2\psi^3)
\label{x}
\end{eqnarray}
\begin{eqnarray}
  b^{++}_r &=&
  +i\oint dz e^{-\frac{\phi}{2}}
   [\gamma^{r-\frac{1}{2}}S^{+++-+}+\gamma^{r+\frac{1}{2}}S^{-+--+}]    
             \nonumber\\
  b^{+-}_r &=&
  +i\oint dz e^{-\frac{\phi}{2}}
   [\gamma^{r-\frac{1}{2}}S^{++++-}+\gamma^{r+\frac{1}{2}}S^{-+-+-}] 
      \nonumber \\
  b^{-+}_r &=&
  -i\oint dz e^{-\frac{\phi}{2}}
   [\gamma^{r-\frac{1}{2}}S^{+---+}+\gamma^{r+\frac{1}{2}}S^{--+-+}]
      \nonumber \\
  b^{--}_r &=&
  -i\oint dz e^{-\frac{\phi}{2}}
   [\gamma^{r-\frac{1}{2}}S^{+--+-}+\gamma^{r+\frac{1}{2}}S^{--++-}] .
\label{b}
\end{eqnarray}
The expressions \eqn{a} are already given in \cite{9806194},
and \eqn{x}, \eqn{b} are our original results.

One can verify the following commutation relations 
\begin{eqnarray}
  \left[ a_m^{AK}\;,\; a_n^{BL} \right]
  &=& -kpm\delta_{m+n}\epsilon^{AB}\epsilon^{KL}  \nonumber \\
  \left[ x_m^{AK}\;,\; a_n^{BL} \right]
  &=&  kp \delta_{m+n}\epsilon^{AB}\epsilon^{KL}  \nonumber \\
  \left[ x_m^{AK}\;,\; x_n^{BL} \right] &=& 0    \nonumber \\
  \left\{ b_r^{\alpha K}\;,\; b_s^{\beta L} \right\} ,
  &=& kp\delta_{r+s}\epsilon^{\alpha\beta}\epsilon^{KL}
\label{comm}
\end{eqnarray}
where $\epsilon^{+-}=-\epsilon^{-+}=1$ for any kind of indices.
They also ``correctly" act on our space-time vacuum:
\bea
 a^{AK}_n \vac & = & 0  , ~~~ (\forall n \geq 0) \nonumber \\
 b^{\alpha K}_r \vac & =& 0 , ~~~ (\forall r \geq 1/2)  
\label{fock}
\eea  

Furthermore we can prove by straightforward (but, the parts including
the OPEs with respect to spin fields are rather complicated) calculations
that they have the following commutation relations with the SCA
generators:
\begin{eqnarray}
  \left[ {\cal L}_m           \;,\; a_n^{AL} \right] 
  &=& -n a_{m+n}^{AL}  \nonumber\\
  \left[ {\cal K}_m^a         \;,\; a_n^{AL} \right]
  &=& 0                \nonumber\\
  \left[ {\cal G}_r^{\alpha A}\;,\; a_n^{BK} \right]
  &=& \epsilon^{AB}b_{n+r}^{\alpha K} \nonumber\\
  \left[ {\cal L}_m           \;,\; x_n^{AL} \right] 
  &=& -n x_{m+n}^{AL} \nonumber\\
  \left[ {\cal K}_m^a         \;,\; x_n^{AL} \right]
  &=& 0 \nonumber\\
  \left[ {\cal G}_r^{\alpha A}\;,\; x_n^{BK} \right]
  &=& 0 \nonumber\\
  \left[ {\cal L}_m              \;,\; b_s^{\alpha K} \right]
  &=& -(\frac{m}{2}+s)b_{m+s}^{\alpha K}  \nonumber \\
  \left[ {\cal K}_m^a\;,\; b_s^{\alpha K} \right]
  &=& \frac{1}{2}\left(\sigma^a\right)^\alpha_{\;\beta}b_{m+s}^{\beta K}
  \;\;\;\;\; ; \;\;\;
  \sigma^a = \left(\begin{array}{cc} 
             (\sigma^a)^-_{\;-} & (\sigma^a)^-_{\;+}  \\
             (\sigma^a)^+_{\;-} & (\sigma^a)^+_{\;+}  
                   \end{array}\right)  \nonumber       \\
  \left\{{\cal G}_r^{\alpha A}   \;,\; b_s^{\beta K} \right\}
  &=& \epsilon^{\alpha\beta}a_{r+s}^{AK}
\end{eqnarray}
  In this way  we have found that we can safely identify 
$\left\{ a^{AK}_n, x^{AK}_0, b^{\alpha K}_r \right\}$ 
with  the mode operators of SCFT on diagonal $T^4$.
  As for $x^{AK}_n$ with $n \ne 0$, we have still no idea on their role in the
spacetime CFT. We hope to discuss elsewhere 
more about them.

One can also construct more general oscillators
in a trivial manner, at least on the Hilbert space of untwisted 
sector. Let ${\cal A}$ be any of $a^{AK}_n, x^{AK}_0, b^{\alpha K}_r $.
We can take ${\cal A}^{(i)}$ $(i=1,\ldots,p)$
 (which was already defined 
as the restriction of ${\cal A}$ on the $i$-th factor of
our untwisted sector Hilbert space) as the desired mode operator.
It is obvious that such oscillators satisfy  the same 
(anti-)commutation relations as those of $T^{4p}/S_p$ SCFT with 
$c=6p$.

Let us consider the case $k=1$. In the space-time CFT,
we have got  the bosonic and fermionic  mode operators 
which have the same algebraic structure as those of fundamental 
fields of $T^{4p}/S_p$ SCFT. From these oscillators,
we can further construct a $N=4$ SCA in the well-known quadratic forms 
of oscillators. Now, a natural question arises: Is this SCA the same one
as the SCA defined in section 2?

Let ${\cal G}_{\msc{GKS}}$ be
any SCA generator introduced in section 2,
and ${\cal G}_{\msc{quad}}$ be
the corresponding operator defined in the quadratic form by the oscillators.
Our question is whether or not 
${\cal G}_{\msc{GKS}} \ket{\alpha} = {\cal G}_{\msc{quad}} \ket{\alpha}$
holds for any state $\ket{\alpha}$ in our space-time Hilbert space.

We already know that any ${\cal G}_{\msc{GKS}} $ and  
${\cal G}_{\msc{quad}}$ have the same commutaiton relations with any of
our oscillators. So, the operator
${\cal G}_{\msc{GKS}} -{\cal G}_{\msc{quad}} $ commutes with all 
of the oscillators. It immediately follows from this fact that 
${\cal G}_{\msc{GKS}} \ket{\alpha} = {\cal G}_{\msc{quad}} \ket{\alpha}$
holds {\em up to null states} as  far as our Hilbert space can be 
spanned by these oscillators. In a few simple examples,
we actually face the situations    
that the differences between them do {\em not} vanish.  
But even in these cases we can observe by direct calculation 
that they become truly spurious states up to some BRST exact terms.

\subsection{Chiral Primaries}
  Now we study the spectrum of chiral primaries of the space-time CFT
and compare it with that of the $T^{4kp}/S^{kp}$-SCFT.
We here set $p=1$ for the time being.

It is a famous fact about the symmetric orbifold of this type
 that there is a sequence of chiral primaries 
\[ \omega^{q,\bar{q}}_{(j)}(z,\bar{z})= 
   \omega^q_{(j)}(z)\bar{\omega}^{\bar{q}}_{(j)}(\bar{z}) \;\;\; ; \;\;\;
   j=0,\frac{1}{2},1,\ldots \] for each element $\omega^{q,\bar{q}}$ of cohomology of $T^4$, and this operator
has the R-charge $\dsp (Q,\bar{Q}) = (\frac{q}{2}+j, \frac{\bar{q}}{2}+j)$.
  The above kind of chiral primaries will be  called as those of 
``single-particle type".
  In addition to these one can obtain many other chiral primaries of
``multi-particle type" by taking products of single-particle ones.
  It is interesting to consider if we can find the same spectrum of
chiral primaries in the space-time CFT.

  Let us begin with the chiral primaries with $j=0$. 
  One can find out the following correspondence:
\begin{eqnarray}
  (q=0) &\longleftrightarrow& 
  \left(\omega^0_{(0)}\right)_n= \delta_{n,0} \nonumber \\
  (q=1) &\longleftrightarrow&  
  \left(\omega^{1,\pm}_{(0)}\right)_r= b_r^{+\pm} \nonumber \\
  (q=2) &\longleftrightarrow& 
  \left(\omega^2_{(0)}\right)_n= {\cal K}_n^+  ,  \label{j0mode}
\end{eqnarray}
or equivelently, we can also express them as the chiral primary {\em states\/};
\bea
\ket{\omega^0_{(0)}} &=& \gamma^{-1}\psi^+ e^{-\phi}(0) \vac  \nonumber\\
\ket{\omega^{1,\pm}_{(0)}} &=& b^{+\pm}_{-1/2}\vac 
  \equiv i \gamma^{-1} S^{+++\mp\pm} e^{-\frac{\phi}{2}}(0) \vac \nonumber\\
\ket{\omega^2_{(0)}} &=& {\cal K}^+_{-1} \vac \equiv \gamma^{-1}\chi^+
   e^{-\phi}(0) \vac ,   \label{j0state} 
\eea
where readers should understand the product such as ``$\gamma^{*}\vac$"
in the sense of the normal product. (Remember  $\vac$ includes
$e^{iY}$ in the definition \eqn{vac}.)
Of course, these states are BRST-invariant and 
satisfy the following relations characterizing the chiral
primary states;
\bea
Q_{\msc{BRST}}\ket{\omega^q_{(0)}} &=& 0  \\
{\cal L}_n \ket{\omega^q_{(0)}} &=& 0 ~~~~(\forall n \geq 1) \nonumber \\
{\cal K}^a_n \ket{\omega^q_{(0)}} &=& 0 ~~~~(\forall n \geq 1) \nonumber \\
{\cal G}^{+A}_r \ket{\omega^q_{(0)}} &=& 0 ~~~~(\forall n \geq -\frac{1}{2})
        \nonumber \\
{\cal G}^{-A}_r \ket{\omega^q_{(0)}} &=& 0 ~~~~(\forall n \geq \frac{1}{2}) .
\eea
We also obtain 
\be
{\cal L}_0 \ket{\omega^q_{(0)}} = {\cal K}^3_0 \ket{\omega^q} = \frac{q}{2}\ket{\omega^q_{(0)}} .
\ee
We shall call it the ``chiral primary state with R-charge $\dsp 
\frac{q}{2}$". 


It is clear that the chiral primaries with $j=0$ \eqn{j0mode}, 
\eqn{j0state} should be identified with those within the untwisted sector
of $T^{4kp}/S^{kp}$-SCFT.

  To write down the chiral primaries of single-particle type for generic $j$
one needs some preparations.
  Let us define the following operators on the worldsheet  
\begin{eqnarray}
  V_{j,m} &\equiv&
  \gamma^{j+m}\exp\left[\frac{2j\varphi}{\alpha_+}\right] \\
  \tilde{V}_{j,m} &\equiv&
  \frac{1}{\sqrt{(j+m)!(j-m)!}}\tilde{\gamma}^{j+m}
  \exp\left[\frac{-2ij\tilde{\varphi}}{\alpha_+}\right]
\end{eqnarray}
which have the following OPE's with $SL(2,R)$ and $SU(2)$ currents
\begin{eqnarray}
  j^3(z)  V_{j,m}(w) &\sim& \frac{m}{z-w}V_{j,m}(w) \nonumber \\
  j^\pm(z)V_{j,m}(w) &\sim& \frac{m\mp j}{z-w}V_{j,m\pm 1}(w) \nonumber \\
  k^3(z)  \tilde{V}_{j,m}(w) &\sim& \frac{m}{z-w}\tilde{V}_{j,m}(w) \nonumber\\
  k^\pm(z)\tilde{V}_{j,m}(w) &\sim&
  \frac{1}{z-w}\sqrt{(j\mp m)(j\pm m+1)}\tilde{V}_{j,m\pm 1}(w) .
\end{eqnarray}
Then we can find a chiral primary for each $q$, having the following operators
as modes:
\begin{eqnarray}
  q=0 &\longleftrightarrow&
  \left(\omega^0_{(j)}\right)_n =
  \oint dz e^{-\phi}\tilde{V}_{j,j}V_{j,n}
  \left(\gamma\psi^-+\gamma^{-1}\psi^+-2\psi^3\right) \nonumber \\
  q=1 &\longleftrightarrow&
  \left(\omega^{1,\pm}_{(j)}\right)_r =
  \oint dz e^{-\frac{\phi}{2}}\tilde{V}_{j,j}\left[
  V_{j,r-\frac{1}{2}}S^{+++\mp\pm} + 
 V_{j,r+\frac{1}{2}}S^{-+-\mp\pm}\right] \nonumber \\
  q=2 &\longleftrightarrow&
  \left(\omega^2_{(j)}\right)_n =
  \oint dz e^{-\phi}\tilde{V}_{j,j}V_{j,n}\chi^+ \label{jmode}
\end{eqnarray}
We can also express them as the chiral primary states;
\bea
\ket{\omega^0_{(j)}} &=& e^{-\phi} \psi^+ 
       \tilde{V}_{j,j} V_{j,-j-1}(0) \vac  \nonumber\\
\ket{\omega^{1,\pm}_{(j)}} &=& i  e^{-\frac{\phi}{2}}
        S^{+++\mp\pm} \tilde{V}_{j,j} V_{j,-j-1}(0) \vac \nonumber\\
\ket{\omega^2_{(j)}} &=&  e^{-\phi} \chi^+
      \tilde{V}_{j,j} V_{j,-j-1}(0) \vac ,  \label{jstate}   
\eea

There are some comments here:
First, in tensoring the chiral primaries of left and right movers 
one has to set the quantum number $j$  
for the both movers  equal.
This is because one can think of $j$ as the momentum of $\varphi$, which
parameterizes the radial direction of $AdS_3$ and is evidently
non-compact.
This aspect about the R-charge is consistent with the known results
about the spectrum of cohomology of symmetric orbifolds.

Second, let us discuss about the unitarity bound for the R-charge of chiral
primaries. In the above constuction of chiral primaries, the value of 
the quantum number $j$ should have an upper-bound $(k-2)/2\sim k/2$ 
\cite{9806024,9806194}. On the other hand, 
it is known that the unitarity bound for
the chiral primaries of the {\em single\/} pariticle type in the $T^{4kp}/S^{kp}$-SCFT is equal to $\sim kp/2$\footnote
     {The unitarity bound for general 
      (namely, including multi-particle types) chiral primary is 
         equal to $\dsp \frac{c}{6} = kp$. (Remark that the unit
     of R-charge we used here is $1/2$ of the usual one in $N=2$ SCFT.)
       This is usually called ``stringy exclusion principle" 
      \cite{9804085}. But, it should be emphasized that the unitarity
      bound for the {\em single\/} particle type is $1/2$ of it.}
As was already claimed by several authors \cite{9806024,9806194,9812027},
in the case of $p=1$
both  of these results are consisitent. But, how about the case $p>1$?
In \cite{9806194}, it was discussed that the missing states with 
$k/2 \leq j \leq pk/2$ might be multi-particle states.
However, this interpretation seems to contradict 
with the Kaluza-Klein spectrum in SUGRA 
discussed in \cite{9806104}.
We do need  the chiral primaries of {\em single}-particle type
possessing large R-charges beyond $k/2$. 

In the previous section we proposed the Hilbert space of 
space-time CFT for general $p$. Especially, we discussed 
a phenomenon that the levels of $SU(2)$ and $SL(2,\br)$ current algebras 
effectively enhance to $pk$ 
on the worldsheet of the maximally joined string.
So, we can get  $ pk/2$ instead of $k/2$ as the unitarity 
bound for the chiral primaries of single particle type, which
is no other than the result we want! We can find out the 
single-particle chiral primary saturating the bound $ pk/2$
on this worldsheet without any problem.

~

  Let us then turn to the multi-particle type.
They should be defined as some products of chiral primaries of 
single-particle type.  
We would like to discuss here the chiral primaries of multi-particle
type which are obtained by multiplying those of single particle type 
with $j=0$.
  The following discussion can be easily generalized to the cases
of more complicated multi-particle types.

  By taking the product of $\omega^p_{(0)}(z)$ one can obtain the following
three types of multi-particle chiral primaries:
\begin{equation}
  \left[ \omega^2_{(0)}(z)\right]^i\;\; , \;\;
  \left[ \omega^2_{(0)}(z)\right]^i\omega^{1,\pm}_{(0)}(z) \;\; , \;\;
  \left[ \omega^2_{(0)}(z)\right]^i
  \omega^{1,+}_{(0)}(z)\omega^{1,-}_{(0)}(z)
\label{multi}
\end{equation}
More concretely, we can also express them in terms of mode operators.
For instance, the modes of the first example can be written as 
\begin{equation}
  \left[ \omega^2_{(0)}(z)\right]^i_n 
     = \sum_{n_1,\ldots, n_i} {\cal K}^+_{n_1}\cdots {\cal K}^+_{n_i}\cdot
        \delta_{n,n_1+\ldots+n_i},
\end{equation}
and the corresponding chiral primary states is as follows;
\be
  \ket{[\omega^2_{(0)}(z)]^i} = \left( {\cal K}^+_{-1}\right)^i \, \vac .
\label{omega2i}
\ee 
Here we give also a comment about the spectrum. 
Let us first consider the case of $p=1$.
In this expression \eqn{omega2i},
we have an upper-bound $k$ for the value $i$, because 
$\left( {\cal K}^+_{-1}\right)^{k+1} \, \vac =0 $ (null state) holds.  

On the other hand, the corresponding operators in the $T^{4kp}/S_{kp}$
SCFT, can be expressed in the simple form of products of fermionic 
oscillators. (Recall that $\omega^2_{(0)}$ corresponds to 
the operator of the form 
$\dsp \sim \sum_{A=1}^{kp} \Psi^{++}_A \Psi^{+-}_A$.) 
Hence we can obtain the bound $i\leq kp$ which is due to nothing but
the fermi-statistics. We again find both are consisitent, if $p=1$.

To treat  the cases with $p>1$, 
let us consider the untwisted sector of  space-time Hilbert space, 
which is nothing  but a symmetrized tensor product of single-string Hilbert 
space. We can immediately discover the desired multi-particle
chiral primary saturating the unitarity bound $kp$ in this Hilbert space. 
In fact, we only have to consider $p$-symmetrized tensor product of 
$\left(\omega^2_{(0)}\right)^k$. (In other words, one can simply imagine 
the situation that $k$-particle chiral primary state lives on  each  
worldsheet of $p$ short strings, which gives a $kp$-particle chiral
primary state.) We can also present the similar arguments for 
the second and third examples of multi-particle chiral primaries \eqn{multi}.

In this way, we have obtained a satisfactory correspondence 
between the chiral primaries in the space-time CFT and those 
of $T^{4kp}/S_{kp}$ SCFT.

\bigskip

\section{Conclusions and Discussions}
\cleqn 

In this paper, according to the framework presented
in \cite{9806194}, we investigated the space-time CFT
associated with the superstring on the background 
$AdS_3 \times S^3 \times T^4$, 
which is an interesting 
proposal of boundary CFT in the $AdS_3/CFT_2$-correspondence.  
We explicitly constructed the suitable vacuum of the Hilbert space
of the space-time CFT in the case $p=1$. 
In the general cases with 
$p>1$, we proposed that
\be
{\cal H}_{\msc{space-time}} = \bigoplus_{\sn \in Y_p} \,
   \left[ \bigotimes_{i=1}^{l(\sn)} \, 
{\cal H }_k ^{\sbz_{n_i}}\right]^{\Gamma_{\sn}}.
\label{sthilb}
\ee
In this expression, $\n \equiv (n_1, \ldots, n_l) \in Y_p$ means 
any partition of $p$ (namely, $Y_p$ denotes the set of Young tableaus
composed of $p$ boxes), $l\equiv l(\n)$ 
expresses the ``depth" of tableau $\n$.
$\Gamma_{\sn}$ is the symmetry group composed of any permutations of 
twisted strings with the same length as before. 
${\cal H}_k$ expresses the Hilbert space associated to a single
$AdS_3$-string with level $k$ for the WZW sectors, 
in which we first constructed
the space-time vacuum in section 3. We also discussed about an effective
enhancement of the level on the worldsheet of twisted strings, and 
we observed in section 4, how this enhancement of level can resolve
the problems about the chiral primaries. 
In order to confirm our proposal about space-time Hilbert space,
it may be significant to formulate carefully the Matrix string 
theory on $AdS_3$-background, which will become our most important task in
future. 

At this stage 
we would like again to make some comments on the compatibilities
of the work of GKS and the work 
\cite{9812046}.
In the framework given in \cite{9806194,9812046}
the fundamental  claim  of $AdS_3/CFT_2$-duality may be encoded
somewhat formally  as follows;
 \begin{eqnarray}
  \lefteqn{
  \left<{\cal O}_1(\varphi,\beta,\gamma; x_1)
	{\cal O}_2(\varphi,\beta,\gamma; x_2)\cdots
	{\cal O}_n(\varphi,\beta,\gamma; x_n) 
  \right>_{{\rm String~on~}AdS_3}
  } ~~~~~~~~~~\nonumber \\
  &=& \left<{\rm vac}|{\rm T}\left[
	{\cal O}_1(\varphi,\beta,\gamma; x_1)
	{\cal O}_2(\varphi,\beta,\gamma; x_2)\cdots
	{\cal O}_n(\varphi,\beta,\gamma; x_n)\right]
   |{\rm vac}\right>,			
\label{correlator}
\end{eqnarray}
where ${\cal O}_i(\varphi,\beta,\gamma; x)$ stands for 
some chiral primary operator of the boundary theory inserted at the point
$x \in \mbox{boundary}$\footnote
    {Although in \cite{9806194} and this paper only the expressions of 
     mode oscillators of chiral primaries are given,  
      we can always define this operator
     $O_i(x)$ at least in a formal sense.}. 
One must remember that 
${\cal O}_i(\varphi,\beta,\gamma; x)$
is made up of the degrees of freedom of string theory in bulk.
So, one can calculate these correlation functions in two ways,
{\em as string theory in bulk and as the space-time CFT\/}.
Namely, the left hand side of the above expression \eqn{correlator}
should be evaluated   in the string theory on $AdS_3$ back-ground.
This  includes the integral over the moduli of worldsheets (quantization of 
two-dimensional gravity) and we must sum over all the different topology
of worldsheets, including the non-connected ones.

On the other hand, in the right hand side
the correlator  among  the {\em same\/} operators ${\cal O}_i$ 
must be computated in terms of the operator algebra in the space-time CFT,
which is derived from the fundamental OPE's of 
$\varphi, \beta,\gamma,\ldots$, and our definition of space-time vacuum
$\vac$. ``T'' denotes the 
T-order (radial order) defined  with respect to the boundary points 
$x_1,\, \ldots ,\, x_n$.
Here we again emphasize that the space-time CFT is {\em not\/} a string 
theory {\em but\/}  
a two-dimensional CFT defined on the {\em fixed\/} worldsheet 
(that is, the boundary of $AdS$-space), although
it is made up of the fundamental field contents of 
the string theory in bulk. 
For the calculation in the right hand side
we need not integrate moduli of worldsheet
and not take the summation with respect to
the topology of worldsheets. 
  
One can say that the analyses in \cite{9812046} are concerned with 
the left hand side of \eqn{correlator} and 
our work and that of \cite{9806194} are studies about 
the right hand side. Both of them may be  compatible in this sense.
Of course, in order to justify this compatibility completely, one must 
prove the identity \eqn{correlator} for arbitrary chiral primaries, 
at least in the semi-classical level of bulk theory.
It will be a very important and challenging problem in our future works.
In any case  we believe that 
studies  along the lines of \cite{9806194} and \cite{9812046}  
will play some complementary roles to each other 
in understanding  of the $AdS_3/CFT_2$-duality. 

There are also  some comments about the other open problems.
First, it is an important task
to analyse more about the physical spectrum (BRST cohomology)
of $AdS_3$-string theory, at least, for the chiral primaries 
(in the sense of space-time, of course). We here only point out
the following:
Because the Virasoro generators of the world-sheet and space-time
CFTs are different, the meanings of ``primary field" for both theories
are different in general. So, our analysis in section 4 is not
complete, since we merely analyzed   the states which are primary
not only in the sense of space-time 
but also in the sense of world-sheet.

Second, we must remark the fact that the space-time conformal 
algebra proposed by GKS does {\em not\/} commute with the screening charges 
in the $SL(2 ,\br)$-sector (especially, for the higher modes).
Therefore, precisely speaking, we may have to regard $AdS_3$-string 
as the Wakimoto free field  system itself, instead of $SL(2,\br)$-WZW model.
(Namely, our claim here is that we need not take the 
Bernard-Felder cohomology  \cite{BF} to define the physical Hilbert space.)
If so, one might have to reconsider about the unitarity of the string
theory on $AdS_3$, since the no-ghost theorem given in \cite{9806024}
was proved under the assumption that the Hilbert space of 
$SL(2,\br)$-sector is an irreducible module of current algebra, not
the Fock module of Wakimoto free fields.

As the third comment, we would like to mention on the similarity of our 
space-time Hilbert space with that of $T^{4kp}/S_{kp}$.   
This can be written as
\be
{\cal H}_{T^{4kp}/S_{kp}} = \bigoplus_{\sn \in Y_{kp}} \,
   \left[ \bigotimes_{i=1}^{l(\sn)} \, 
{\cal H }_{T^4} ^{\sbz_{n_i}}\right]^{\Gamma_{\sn}},
\label{symhilb}
\ee
where ${\cal H }_{T^4}$ denotes the Hilbert space of $N=4$ SUSY
$\sigma$-model on $T^4$ with $c=6$.
In fact, in the case $k=1$, this has the same structure
as \eqn{sthilb}. However, in the general cases with $k>1$, 
it seems that we have a discrepancy in these structures.
Very recently, in \cite{9812027} it is also claimed that 
the spectra of massive modes with KK momenta and windings along $T^4$ 
in  the space-time CFT does not coincide with 
those of $T^{4kp}/S_{kp}$-SCFT on every points
of moduli space {\em except the case $k=1$}. 
The observation here might have some relation
to this claim. We wish to further discuss this problem elsewhere.

Our work in this paper was only concerned with the formulation.
We would also like to apply our formalism to
 more ``physical" problems,
namely, calculations of correlation functions and their applications to
BTZ black-hole physics\cite{9204099}. 
Especially, it may be interesting to generalize 
our results to superstring on a BTZ black-hole background. 
To this aim we will have to formulate
the space-time CFT on torus. If we succeed in it completely, 
we will be able to define the quantum theory 
of BTZ black-holes with finite temperature
{\em from the stringy viewpoint}, and it will become 
a significant subject to compare it with the previous works  
on BTZ black-holes based on 3-dimensional gravity  \cite{btz}.

~

~

\noindent{\bf Note Added:}
After completing the essential part of this work, we became aware 
of the paper \cite{9812027}, which has some overlap with our results
in section 4.

~

\noindent{\large \bf Acknowledgment} \\
  We would like to thank T. Eguchi, I. Bars and J. de Boer
  for discussions and useful comments.
  We should acknowledge University of Southern California
 for kind hospitality. Part of this work was done 
during our visit there.

K.H is supported in part by JSPS Research Fellowships,
 and also Y.S is supported in part by the Gran-in-Aid  from
the Ministry of Education, Science and Culture, Priority Area:
``Supersymmetry and Unified Theory of Elementary Particles"
$(\sharp 707)$

\newpage
\noindent
{\Large \bf Appendix}
\appendix

\section{Fermions, Spin Fields and their OPE's}
  Here we shall give a definition of spin fields by their OPE's.
  Firstly we define a set of spinors 
$\psi^\mu (\mu=\pm 1, \pm 2, \pm 3, \pm 4, \pm 5)$
by the following equations:
\begin{eqnarray}
 \psi^{\pm 1} &\equiv& \psi^1\pm i\psi^2 \nonumber\\
 \psi^{\pm 2} &\equiv& \chi^1\pm i\chi^2 \nonumber\\
 \psi^{\pm 3} &\equiv& \chi^3 \pm \psi^3 \nonumber\\
 \psi^{\pm 4} &\equiv& \lambda^1 \pm i\lambda^2 \nonumber\\
 \psi^{\pm 5} &\equiv& \lambda^3 \pm i\lambda^4
\end{eqnarray}
They have the following OPE:
\begin{eqnarray}
 \psi^\mu(z)\psi^\nu(w) 
              &\sim  & \frac{2\eta^{\mu\nu}}{z-w} \\
 \eta^{\mu\nu}&\equiv& \delta^{\mu + \nu}
\end{eqnarray}

  Secondly we introduce the spin fields $S^A(z)$ as the fields satisfying 
the following OPE's
\begin{equation}
  \frac{1}{2}:\psi^\mu\psi^\nu:(z)S^A(w) \sim
  -\frac{1}{z-w}(\Gamma^{\mu\nu})^A_{\;\;B}S^B(w)
\label{OPE1}
\end{equation}
where $\Gamma^{\mu\nu}\equiv \Gamma^{[\mu}\Gamma^{\nu]}$ and $\Gamma^\mu$
is the ten-dimensional gamma-matrix satisfying 
\begin{equation}
  \left\{ \Gamma^\mu, \Gamma^\nu\right\} = \eta^{\mu\nu}
\end{equation}
  Ten dimensional gamma-matrix is of course 32-dimensional, and we use as
spinorial indices ($A,B,\ldots$) the sets of five signs as in the main text,
namely, $S^A$ consists of 32 components
\[ S^{+++++}\;,\; S^{++++-}\;,\; \ldots\;,\; S^{-----}. \]
  As a matter of fact, half of the above 32 components play no role.
Throughout this paper we restrict ourselves to the components of spin fields
with the fixed chirality, namely, 
$S^{\epsilon_1\epsilon_2\epsilon_3\epsilon_4\epsilon_5}$ with
$\Pi\epsilon_i= -1$.
  The OPE's of spin fields themselves are
\begin{equation}
  S^A(z)S^B(w)\sim (z-w)^{-\frac{3}{4}}(\Gamma_\mu C)^{AB}\, 
    \sqrt{\frac{k}{2}}\psi^\mu
\label{OPE2}
\end{equation}
where $C$ is the charge conjugation matrix.
 
  Equations (\ref{OPE1}) and (\ref{OPE2}) are enough for us to prove all the
equations in this paper, if the gamma matrices and the charge conjugation
matrix have the following components:
\begin{eqnarray}
  \Gamma_{\pm 1}=\Gamma^{\mp 1} &=&
    -\sigma_\pm \otimes \sigma_3 \otimes 1 \otimes \sigma_3 \otimes 1 
        \nonumber\\
  \Gamma_{\pm 2}=\Gamma^{\mp 2} &=&
    -1 \otimes \sigma_\pm \otimes \sigma_3 \otimes \sigma_3 \otimes 1 
       \nonumber\\
  \Gamma_{\pm 3}=\Gamma^{\mp 3} &=&
     \sigma_3 \otimes 1 \otimes \sigma_\pm \otimes \sigma_3 \otimes 1 
      \nonumber \\
  \Gamma_{\pm 4}=\Gamma^{\mp 4} &=&
     1 \otimes 1 \otimes 1 \otimes \sigma_\pm \otimes 1 
       \nonumber \\
  \Gamma_{\pm 5}=\Gamma^{\mp 5} &=&
     \sigma_3\otimes\sigma_3\otimes\sigma_3\otimes\sigma_3\otimes\sigma_\pm
       \nonumber\\
  C &=& \epsilon\otimes\epsilon\otimes\epsilon\otimes\epsilon\otimes\epsilon 
\end{eqnarray}
with
\begin{eqnarray}
  \sigma^a &=&
  \left(\begin{array}{cc} (\sigma^a)^+_{\;\;+} & (\sigma^a)^+_{\;\;-} \\
                          (\sigma^a)^-_{\;\;+} & (\sigma^a)^-_{\;\;-}
  \end{array}\right)  \nonumber \\
  \sigma^+ &=&
  \left(\begin{array}{cc} 0 & 1 \\ 0 & 0 \end{array}\right)\nonumber \\
  \sigma^- &=&
  \left(\begin{array}{cc} 0 & 0 \\ 1 & 0 \end{array}\right) \nonumber \\
  \sigma^3 &=&
  \left(\begin{array}{cc} 1 & 0 \\ 0 &-1 \end{array}\right) \nonumber \\
  \epsilon &=&
  \left(\begin{array}{cc} \epsilon^{++} & \epsilon^{+-} \\
  \epsilon^{-+} & \epsilon^{--} \end{array}\right)
  =\left(\begin{array}{cc} 0 & 1 \\-1 & 0 \end{array}\right)
\end{eqnarray}
or  equivalently, 
\begin{eqnarray}
  (\Gamma_{\pm 1})^{\pm \epsilon_2 \epsilon_3 \epsilon_4 \epsilon_5}
                  _{\mp \epsilon_2 \epsilon_3 \epsilon_4 \epsilon_5}
    &=& -\epsilon_2\epsilon_4  \nonumber \\
  (\Gamma_{\pm 2})^{\epsilon_1 \pm \epsilon_3 \epsilon_4 \epsilon_5}
                  _{\epsilon_1 \mp \epsilon_3 \epsilon_4 \epsilon_5}
    &=& -\epsilon_3\epsilon_4 \nonumber \\
  (\Gamma_{\pm 3})^{\epsilon_1 \epsilon_2 \pm \epsilon_4 \epsilon_5}
                  _{\epsilon_1 \epsilon_2 \mp \epsilon_4 \epsilon_5}
    &=& \epsilon_1\epsilon_4 \nonumber \\
  (\Gamma_{\pm 4})^{\epsilon_1 \epsilon_2 \epsilon_3 \pm \epsilon_5}
                  _{\epsilon_1 \epsilon_2 \epsilon_3 \mp \epsilon_5}
    &=& 1 \nonumber \\
  (\Gamma_{\pm 5})^{\epsilon_1 \epsilon_2 \epsilon_3 \epsilon_4 \pm}
                  _{\epsilon_1 \epsilon_2 \epsilon_3 \epsilon_4 \mp}
    &=& \epsilon_1\epsilon_2\epsilon_3\epsilon_4 \nonumber \\
  C^{\epsilon_1 \epsilon_2 \epsilon_3 \epsilon_4 \epsilon_5,
     \tau_1 \tau_2 \tau_3 \tau_4 \tau_5} &=&
        \epsilon^{\epsilon_1\tau_1}\epsilon^{\epsilon_2\tau_2}
        \epsilon^{\epsilon_3\tau_3}\epsilon^{\epsilon_4\tau_4}
        \epsilon^{\epsilon_5\tau_5}
\end{eqnarray}



\newpage

\begin{thebibliography}{99}

\bibitem{9806194}
A. Giveon, D. Kutasov, N. Seiberg,
{\it ``Comments on String Theory on $AdS_3$''},
hep-th/9806194.

\bibitem{9711200}
J. Maldacena,
{\it ``The Large $N$ Limit of Superconformal Field Theories
and Supergravity''},
Adv. Theor. Math. Phys. {\bf 2} (1998) 231-252,
hep-th/9711200.

\bibitem{9802150}
E. Witten,
{\it ``Anti De Sitter Space And Holography''},
Adv. Theor. Math. Phys. {\bf 2} (1998) 253-291,
hep-th/9802150.

\bibitem{9802109}
S. Gubser, I. Klebanov, A. Polyakov,
{\it ``Gauge Theory Correlators from Non-Critical String Theory''},
Phys. Lett. {\bf B428} (1998) 105-114,
hep-th/9802109.

\bibitem{Brown}
J. Brown, M. Henneaux, 
{\it ``Central Charges In The Canonical Realization Of Asymptotic Symmetries:
An Example from Three-Dimensional Gravity''},
Commun. Math. Phys. {\bf 104} (1986) 207.

\bibitem{9712251}
A. Strominger,
{\it ``Black Hole Entropy from Near-Horizon Microstates''},
J. High Energy Phys. {\bf 02} (1998) 009,
hep-th/9712251.

\bibitem{9802076}
M. Banados, T. Brotz, M. Ortiz,
{\it ``Boundary dynamics and the statistical mechanics of
the 2+1 dimensional black hole''},
hep-th/9802076; 
M. Banados,
{\it ``Global Charges in Chern-Simons theory and the $2+1$ black hole''},
Phys. Rev. {\bf D52} (1996) 5816,
hep-th/9405171.

\bibitem{9804085}
J. Maldacena, A. Strominger,
{\it ``$AdS_3$ Black Holes and a Stringy Exclusion Principle''},
hep-th/9804085.

\bibitem{9804111}
E. Martinec,
{\it ``Matrix Models of $AdS$ Gravity''},
hep-th/9804111; 
{\it ``Conformal Field Theory, Geometry, and Entropy''},
hep-th/9809021

\bibitem{9804166}
S. Deger, A. Kaya, E. Sezgin, P. Sundell,
{\it ``Spectrum of $D=6, N=4b$ Supergravity on $AdS_3 \times S^3$''},
hep-th/9804166.

\bibitem{9805165}
M. Banados, K. Bautier, O. Coussaert, M. Henneaux, M. Ortiz,
{\it ``Anti-de Sitter/CFT Correspondence in Three-Dimensional Supergravity''},
Phys. Rev. {\bf D58} (1998) 085020,
hep-th/9805165.

\bibitem{9806026}
S. Carlip,
{\it ``What We Don't Know about BTZ Black Hole Entropy''},
Class. Quant. Grav. {\bf 15} (1998) 3609-3625,
hep-th/9806026.

\bibitem{9811002}
K. Ito,
{\it ``Extended Superconformal Algebras on $AdS_3$''},
hep-th/9811002.

\bibitem{9811245}
S. Elitzur, O. Feinerman, A. Giveon, D. Tsabar,
{\it ``String Theory on $AdS_3 \times S^3 \times S^3 \times S^1$''},
hep-th/9811245.

\bibitem{9812027}
D. Kutasov, F. Larsen, R. Leigh,
{\it ``String Theory in Magnetic Monopole Backgrounds''},
hep-th/9812027.

\bibitem{9812046}
J. de Boer, H. Ooguri, H. Robins, J. Tannenhauser,
{\it ``String Theory on $AdS_3$''},
hep-th/9812046.

\bibitem{CHS}
C. Callan, J. Harvey, A. Strominger,
Nucl. Phys. {\bf B359} (1991) 611;
Nucl. Phys. {\bf B367} (1991) 60.
 
\bibitem{DGHR}
A. Dabholkar, G. Gibbons, J. Harvey, F. Ruiz-Ruiz,
Nucl. Phys. {\bf B340} (1990) 33.
 
\bibitem{9601177}
A. Tseytlin, 
Mod. Phys. Lett. {\bf A11} (1996) 689,
hep-th/9601177.

\bibitem{Wakimoto}
M. Wakimoto,
Commun. Math. Phys. {\bf 104} (1986) 605.

\bibitem{FMS}
D. Friedan, E. Martinec, S. Shenker,
Nucl. Phys. {\bf B271} (1986) 93.

\bibitem{9806024}
J. Evans, M. Gaberdiel, M. Perry,
{\it ``The No-ghost Theorem for $AdS_3$ and the Stringy
Exclusion Principle''},
Nucl. Phys. {\bf B535} (1998) 152-170,
hep-th/9806024.

\bibitem{9703030}
R. Dijkgraaf, E. Verlinde, H. Verlinde,
{\it ``Matrix String Theory''},
Nucl. Phys. {\bf B500} (1997) 43,
hep-th/9703030;
L. Motl, {\it ``Proposals on nonperturbative superstring
 interactions''},
hep-th/9701025.
 
\bibitem{9806104}
J. de Boer,
{\it ``Six-Dimensional Supergravity on $S^3 \times AdS_3$
 and $2$d Conformal Field Theory''},
hep-th/9806104.

\bibitem{BF}
D. Bernard, G. Felder,
Commun. Math. Phys. {\bf 127} (1990) 145.

\bibitem{9204099}
M. Banados, C. Teitelboim, J. Zanelli,
{\it ``The Black Hole in Three Dimensional Space Time''},
Phys. Rev. Lett. {\bf 69} (1992) 1849,
hep-th/9204099.

\bibitem{btz}
E. Teo,
{\it ``Black hole absorption cross-sections and 
the anti-de Sitter -- conformal field theory correspondence''} 
Phys.Lett. {\bf B436} (1998) 269,
hep-th/9805014;
E. Keski-Vakkuri, {\it ``Bulk and boundary dynamics in BTZ black
 holes''}, hep-th/9808037;
U. Danielsson, E. Keski-Vakkuri and M. Kruczenski,
{\it ``Vacua, Propagators, and Holographic Probes in $AdS/CFT$''},
hep-th/9812007.

\end{thebibliography}
\end{document}